# High-Performance Nuclear Demagnetization Refrigerator Reaching 50 $\mu$K

Zhengchao Liu,[a,b†] Junyang Song,[a†] Bingbing Tong,[a] and Hiroki Ikegami[a,*]

[a]*Beijing National Laboratory for Condensed Matter Physics, Institute of Physics, Chinese Academy of Sciences, Beijing, 100190, China.*

[b]*School of Physical Sciences, University of Chinese Academy of Sciences, Beijing, 100049, China.*

[†] These authors contributed equally

[*] Corresponding author.

*E-mail address*: hikegami@iphy.ac.cn (H. Ikegami).

**Abstract**

Research in the deep sub-mK regime around 0.1 mK may open new opportunities in condensed matter physics in light of recent rapid progress in quantum technologies and quantum materials. Here, we report the design and performance of a nuclear demagnetization system that provides efficient access to the deep sub-mK regime for a wide range of experiments, using a standard wet dilution refrigerator with a medium-sized copper nuclear stage. We show that the temperature measured by a Pt NMR thermometer reaches 50 $\mu$K and that the refrigerator can maintain 100 $\mu$K for more than ten days. Remarkably, even with only 12 hours of precooling, the refrigerator cools into the deep sub-mK regime and can maintain 100 $\mu$K for more than three days, providing sufficient time for sample measurements. Numerical analyses indicate that the copper nuclear-spin system is cooled to about 18 $\mu$K, whereas a local heat leak to the thermometer limits the thermometer temperature to about 50 $\mu$K. These features establish the system as a practical platform for efficient experiments in the deep sub-mK regime.

## I. INTRODUCTION

The sub-mK temperature regime has provided access to exotic and fragile quantum phenomena in condensed matter systems. Early studies led to landmark discoveries of superfluidity in $^3$He [1–5] and unusual nuclear-spin ordering in solid $^3$He [1, 5–7]. More recently, the sub-mK environment has emerged as an important platform for investigating exotic superconductors [8–11], fragile fractional quantum Hall systems [12–14], and nanoelectronic devices [15–20]. Its importance has been further highlighted in the context of quantum technologies and quantum materials, as illustrated by improved decoherence environments in quantum circuits cooled below 1 mK [21] and the observation of odd-parity superconductivity in the heavy-fermion metal $YbRh_2Si_2$ [22].

Access to even lower temperatures in the deep sub-mK regime around 0.1 mK may open new frontiers in condensed matter systems, particularly in light of recent rapid progress in quantum technologies and quantum materials. In quantum devices, the substantial suppression of thermal noise would provide insight into the intrinsic mechanisms limiting quantum coherence [21, 23, 24]. In quantum materials, exotic quantum phenomena with extremely small energy scales may emerge through interactions with nuclear spins [25–27] or quantum critical fluctuations [28–31]. Candidate systems for such studies should be ultraclean with minimal imperfections, and may also be found among emerging materials [32–40]. For superfluid $^3$He, important low-energy phenomena remain to be explored despite extensive studies in the deep sub-mK regime, such as surface Majorana fermions near the Dirac point [41–44], the topological phase transition [45, 46], and a truly two-dimensional superfluid $^3$He [47–51].

For studies in the sub-mK regime, adiabatic nuclear demagnetization systems have been developed and widely used [52], including recently developed systems based on dry dilution refrigerators [53–57]. Among them, some can reach temperatures well below 100 $\mu$K and maintain them for long periods (several days or longer). Such systems often employ two-stage nuclear demagnetization [58–60] or a large single copper stage with an effective amount of ~100 mol [61, 62]. While these systems are effective for experiments at such temperatures, they often require a specially designed room, as well as appreciable amounts of manpower and cryogens for setup, cooldown, and operation.

In this article, we present an adiabatic nuclear demagnetization refrigerator with a 35-mol copper nuclear stage, based on a wet dilution refrigerator with moderate cooling power. This system reaches 50 $\mu$K and can maintain a temperature of 100 $\mu$K for more than ten days. It was developed to perform a wide range of experiments in the deep sub-mK regime, such as studies of quantum circuits, novel quantum materials, and low-energy phenomena in superfluid $^3$He. To allow for such diverse applications, the system was designed to provide experimental flexibility, including easy replacement of experimental cells and the simultaneous installation of multiple cells.

Although dry nuclear demagnetization systems have recently become popular, we chose a wet dilution refrigerator because wet systems offer important advantages for certain experiments. In the deep sub-mK regime, the thermal relaxation time of some samples becomes very long [52], often requiring the temperature to be held for extended periods (for more than one day), which remains challenging for dry systems because of vibration-induced heating [53, 54, 63]. A low-vibration environment is also essential for experiments that are sensitive to mechanical disturbances, such as experiments on the free surface of superfluid $^3$He [64–70], including surface Majorana fermions [41, 42, 71–73].

## II. THE ADIABATIC NUCLEAR DEMAGNETIZATION SYSTEM

In a single-shot adiabatic nuclear demagnetization refrigeration system, the nuclear spins in a metallic stage, typically copper, are first polarized in a strong magnetic field by removing the heat of magnetization using a dilution refrigerator. After this precooling, the magnetic field is reduced adiabatically to a small but finite value. Ideally, the nuclear-spin temperature decreases in proportion to the magnetic field, and the electron temperature in the stage follows. In practice, however, heat leaks from the environment cause the electron temperature to become higher and also gradually destroy the nuclear-spin polarization, thereby limiting the achievable temperature. Therefore, reducing the heat leak is a key requirement for reaching the deep sub-mK regime. In addition, minimizing the thermal resistance between components is essential to suppress temperature gradients and ensure efficient cooling of the sample.

Our adiabatic nuclear demagnetization system consists of a wet dilution refrigerator, a superconducting heat switch, a nuclear demagnetization stage made of copper, a 9 T superconducting magnet for magnetizing the stage, and thermometers. In designing the system, special attention was paid to suppressing heating caused by vibrations and rf fields, minimizing the thermal resistance within the microkelvin temperature region, and reducing the thermal resistance between the mixing chamber of the dilution refrigerator and the copper stage to shorten the precooling time.

The dilution refrigerator employed for this system is a standard model with a nominal cooling power of 700 $\mu$W at 120 mK (model MNK 126-700, Leiden Cryogenics BV). Because the original radiation shields and inner vacuum chamber (IVC) were not sufficiently large to accommodate a demagnetization stage and sufficient experimental space, we redesigned and remanufactured them to fit our design of the demagnetization system. The cryostat is suspended from a vibration isolation stage weighing 1600 kg supported by four air springs. The springs sit on a frame made of nonmagnetic stainless steel, which is anchored to the floor. The vibration isolation stage and the air springs form a damped oscillator with resonant frequencies of 3.0 and 2.3 Hz in the two horizontal directions and 3.0 Hz in the vertical direction, providing approximately 20 dB attenuation over the frequency range from ~7 Hz to above ~50 Hz relative to the floor. To prevent rf fields from entering the low-temperature parts, the top of the cryostat is covered with an electromagnetically shielded housing that attenuates rf fields by more than 60 dB in the range of 500 kHz – 1 GHz. The shielded housing is connected to a shielded room built next to the cryostat, where all measurements are performed.

Figure 1 schematically shows the system below the mixing chamber inside the IVC. The experimental space, supported by three rods made of Vespel SP-22, is located below the mixing chamber, and the copper nuclear stage is suspended below it. The experimental space consists of upper and lower plates, which are connected by three pillars. The space is designed to be large enough (98 mm in diameter and 215 mm in length) to accommodate multiple experimental cells (up to five: two cells up to 52 mm in diameter and 100 mm in height, and three cells up to 34 mm in diameter and 100 mm in height), together with thermometers and auxiliary components, while allowing easy replacement of the experimental cells. This multi-cell design is particularly useful for carrying out tests of several components used in the sub-mK region, either simultaneously or together with other main experiments. If necessary, it can accommodate large experimental cells by using multiple experimental sites, as well as a compact magnet for applying magnetic fields of, for example, $\sim$3 T [74] to a sample. These features provide the system with the flexibility required for a wide range of experiments.

The nuclear demagnetization stage is made of copper with a nominal purity of 99.9999% (6N) [75]. Its length and diameter are 360 mm and 48 mm, respectively. The stage has 0.8-mm-wide slits to reduce eddy-current heating [Fig. 1(c)]. The effective amount of copper exposed to the magnetic field is 35 mol, as evaluated by integrating $[B(z)/B_\mathrm{c}]^2$ along the stage, where $B(z)$ is the field profile of the 9 T magnet and $B_\mathrm{c}$ is the field at the magnet center. The copper stage is joined to the lower copper plate of the experimental space by electron-beam welding. To improve the thermal conductance, it is annealed at 850 °C for 72 hours in a vacuum ($< 3\times10^{-6}$ mbar), resulting in a residual resistivity ratio (RRR) of about 5500. This value is larger than the RRR of 1000–3000 obtained in previous studies for 4N or 5N copper materials, which have often been used for nuclear stages and were annealed in an oxygen environment to passivate magnetic impurities such as Fe [58,60–62]. This fact suggests that the influence of magnetic impurities in our 6N copper is smaller from the beginning than in 4N and 5N copper, even after those materials are annealed in an oxygen environment.

The stage is magnetized by a 9 T superconducting magnet with a three-inch bore, located in a liquid helium bath. The magnet is equipped with cancellation coils that are wound in series with the main field coil. These coils reduce the magnetic field at the experimental space to below 3 mT even when the main magnet is magnetized up to 9 T.

The parts of the experimental space (the upper and lower plates and the pillars) are made of annealed, high-purity copper with a nominal purity of 6N (RRR $\gtrsim$ 5000), and their surfaces are gold-plated. They are tightly fastened together using M4 stainless steel screws with a sufficiently large torque ($>$ 5 Nm) [76] to minimize temperature differences between the copper stage and the experimental space, as well as within the experimental space itself. From our estimate based on the RRR values of the components and the torques applied to the screws used to join them, the residual resistance from the center of the stage to the far end of the experimental space is about 10 n$\Omega$. (In this article, we use residual electrical resistance as a measure of thermal resistance according to the Wiedemann–Franz law.)

The copper stage is precooled by the dilution refrigerator through the experimental space and a superconducting heat switch [Fig. 1(b)]. The heat switch consists of high-purity aluminum foils (99.9995% purity) and a small superconducting magnet. The magnetic field is applied parallel to the aluminum foils and perpendicular to the heat path. We chose aluminum as the superconducting material for the heat switch because of its relatively low critical magnetic field (10 mT) and easy availability of high-purity material. However, it is one of the most difficult materials for forming good thermal contact with a counterpart material (copper) by soldering or gold plating owing to its hard surface oxide layer. Poor thermal contact could cause the heat switch to be a bottleneck in precooling the nuclear stage (see Appendix B). To make good thermal contact, we employ diffusion bonding to join the aluminum foils and the copper parts directly together [77, 78]. The diffusion bonding also creates a mechanically strong joint, allowing for easy handling during assembly and reliable contact through thermal cycles. As illustrated in Fig. 1(b), four annealed aluminum foils with a thickness of 0.25 mm are sandwiched alternately between five annealed copper plates (99.999% purity) with a thickness of 1 mm. The interfaces between aluminum and copper are then diffusion-bonded, with a contact area of $5\times8$ mm$^2$ on each side of every aluminum foil. (The details of the diffusion bonding are described in Appendix A.) The residual resistance between the two flanges on either side of the heat switch is 27 nΩ at 4 K, estimated from test samples [78].

Owing to the careful construction described above, the residual resistance between the mixing chamber plate and the center of the copper stage is estimated to be as low as about 60 nΩ, including contributions from the components in the experimental space and their contact resistances. This value is small enough to permit efficient precooling of the stage. Our numerical simulations described in Appendix B show that, for a residual resistance of 100 nΩ, for example, the precooling time required to reach 11.5 mK increases by only 11% compared with the zero-resistance case, using the cooling power at the mixing chamber measured under our typical operating conditions [Fig. 8(b)].

In this study, two thermometers are used to monitor the temperature of the experimental space: a cerium magnesium nitrate (CMN) thermometer [52, 79] and a pulsed Pt NMR thermometer [52, 58, 80]. Above 5 mK, temperature is measured using the CMN thermometer, which is calibrated above 50 mK against a calibrated resistance thermometer. Below 5 mK, the Pt NMR thermometer is employed. The thermometer is made using about 3200 Pt wires, each 25 $\mu$m in diameter, with a purity of 99.999%. One end of the wires is TIG-welded to a 2-mm-thick Pt plate, which is screwed to an annealed Ag support with an RRR of 1000; the support is then screwed to the lower plate of the experimental space. The rf coil consists of copper wire 30 $\mu$m in diameter wound on a bobbin made of Vespel SP-22. The wire is thermally anchored to the silver support. Temperature is measured using a PLM-5 (RV-Elektroniikka Oy PICOWATT, Finland) at a frequency of 173.8 kHz, with smaller tipping angles employed at lower temperatures. The Pt NMR thermometer is calibrated against the CMN thermometer at temperatures above 5 mK. The uncertainty in the temperature measured by the Pt NMR thermometer is within $\pm 5\%$ at the lowest temperatures. This uncertainty arises mainly from the rather small signals in the calibrated temperature region and the uncertainty in the measured relationship between the different tipping angles.

## III. PERFORMANCE

### A. Full characterization of the nuclear demagnetization system

The adiabatic nuclear demagnetization process begins with precooling of the copper stage in a magnetic field. We first apply a magnetic field $B$ of 9 T to the stage, which warms the stage to above 50 mK due to the heat of magnetization. After the field reaches 9 T, the stage is precooled by the dilution refrigerator operated with a still-heater power of 10 mW, which corresponds to a cooling power of $\sim$200 $\mu$W at the mixing chamber plate at 100 mK. After 72 hours of precooling, the temperature of the stage reaches 11.5 mK, as shown in Fig. 2(a). This precooling enhances the nuclear-spin polarization, reducing the entropy of the nuclear-spin system by 8.6%.

To precool the stage efficiently, the thermal resistance between the mixing chamber plate and the stage should be sufficiently small. To evaluate it, we perform numerical simulations using the cooling power measured at the mixing chamber plate (Appendix B) and compare the results with the experimental data. From the precooling time required to reach 11.5 mK in Fig. 2(a) (72 hours), the simulations indicate a residual resistance of 85 nΩ. On the other hand, fitting the entire temperature decrease in Fig. 2(a) to the simulation gives a resistance of 130 nΩ. Because of the slight deviation between the simulation and the experimental data, we assign an uncertainty of about $\pm$30 nΩ to the latter resistance value. Taking these estimates into account, we infer the resistance to be around 100 nΩ. This value is reasonably low, although it is somewhat larger than the value described in Sect. II (~60 nΩ).

After precooling the stage, the heat switch is opened and the demagnetization is started. As exhibited in Fig. 2(b), the temperature decreases as the magnetic field is reduced. During demagnetization, the magnetic field is paused multiple times to check the adiabaticity of the demagnetization process and to measure the relationship between different tipping angles of the Pt NMR thermometer. As shown in Fig. 2(c), the temperature $T$ decreases in proportion to $B$ within 5% from ~2.4 T (~3.4 mK) down to ~110 mT (~150 $\mu$K), indicating that the demagnetization process is nearly adiabatic in this temperature range. [We observed a large heat leak from an unknown source that appeared only during field sweeps above ~2.4 T. This heat leak causes the small deviation from adiabaticity observed above ~2.4 T in Fig. 2(c).] Below ~85 mT (~125 $\mu$K), $T$ deviates from the proportionality line. At the minimum field of $B = 12$ mT, the temperature reaches ~46 $\mu$K. The field is then maintained for 37 hours, during which the temperature increases by only ~5 $\mu$K [Fig. 2(d)].

The heat leak to the stage is one of the main factors determining the performance of the adiabatic nuclear demagnetization cooling. To evaluate the heat leak, we monitor the increase in temperature at a constant magnetic field. Typical behavior is shown in Fig. 3. Here, the temperature measured by the Pt NMR thermometer increases at a rate $\dot{T} = 1.1\times10^{-9}$ K/s. The heat leak can be evaluated from

$$\dot{Q} = C_{\rm n}\dot{T}_{\rm n}, \qquad (1)$$

where $C_{\rm n}$ and $T_{\rm n}$ are the heat capacity and temperature of the nuclear-spin system in the copper stage, respectively. Within the high-temperature approximation,

$$C_{\rm n} = n\Lambda\frac{B^2}{T_{\rm n}^2}, \qquad (2)$$

where $n = 35$ mol is the effective amount of the copper stage and $\Lambda = \lambda_{\rm n}/\mu_0$, with $\lambda_{\rm n}$ and $\mu_0$ being the molar nuclear Curie constant and the vacuum permeability, respectively ($\Lambda = 3.22$ $\mu$JK/T$^2$mol for copper) [52]. In the temperature range shown in Fig. 3, the difference between the temperature measured by the Pt NMR thermometer and $T_{\rm n}$ is negligibly small. Therefore, the temperature measured by the Pt NMR thermometer can be regarded as equal to $T_{\rm n}$. For Fig. 3, $C_{\rm n}$ ~ 0.88 J/K, yielding $\dot{Q}$ ~ 1.0 nW. We monitor $\dot{Q}$ at different magnetic fields and temperatures and find that $\dot{Q}$ is always around 1–2 nW at $B \leq 0.25$ T when the magnetic field is kept constant. We note that an increase in temperature like the one shown in Fig. 3 is observable only when $C_{\rm n}$ becomes small (i.e., small $B/T_{\rm n}$). This heat leak does not decrease even after the refrigerator has been maintained below ∼1 K for 45 days, suggesting that it does not originate from time-dependent internal heat sources [52, 58, 81]. Because of this constant background heat leak, any time-dependent heat leak smaller than ∼1 nW would not be observable, even if present. We note that we do not observe the relatively large time-dependent heat leaks of 1–10 nW that were reported in some previous works after refrigerators had been kept below 1 K for 10–20 days (Figs. 7 and 8 in Ref. [81]). This may be related to the small amount of non-crystalline organic materials used (Vespel SP-22: ∼10 g for the supports of the experimental space; Stycast 1266: <1 g; Stycast 2850FT: <1 g) and to the careful annealing of metal parts, which releases strain stored in the original materials and welded joints and may also remove hydrogen possibly trapped in the original copper materials.

The heat leak gives rise to a difference between $T_\mathrm{n}$ and the electron temperature in the copper stage, $T_\mathrm{e}$. To estimate $T_\mathrm{n}$ and $T_\mathrm{e}$ at the minimum field, we employ the heat flow model depicted in Fig. 4. The total heat leak $\dot{Q}$ entering the region below the heat switch first flows into the electron system of the stage and is subsequently transferred to the nuclear-spin system through the spin–lattice coupling given by [52]

$$\dot{T_\mathrm{n}} = (T_\mathrm{e} - T_\mathrm{n})T_\mathrm{n}/\kappa, \quad (3)$$

where $\kappa = 1.2$ Ks is the Korringa constant for copper [52, 82]. From Eqs. (1) and (3), we have

$$\frac{T_\mathrm{e}}{T_\mathrm{n}} = 1 + \frac{\kappa\dot{Q}}{C_\mathrm{n}T_\mathrm{n}^2} = 1 + \frac{\kappa\dot{Q}}{n\Lambda B^2}. \quad (4)$$

This equation describes the increase in $T_\mathrm{e}$ above $T_\mathrm{n}$ caused by $\dot{Q}$.

Using Eq. (4), we first make a quick estimate of $T_\mathrm{n}$ and $T_\mathrm{e}$ to gain preliminary insight into the observed behavior at the minimum field. (A comprehensive analysis is presented later.) For this purpose, we focus on the behavior when the magnetic field is increased from point C to point D in Fig. 2(d). Assuming that this magnetizing process is adiabatic and that $T_\mathrm{Pt} = T_\mathrm{n}$ at point D, where $T_\mathrm{Pt}$ denotes the temperature measured by the Pt NMR thermometer, the nuclear-spin temperature at point C is estimated to be $T_\mathrm{n} = 27$ $\mu$K from the temperature at point D (240 $\mu$K). Using this value in Eq. (4), we then obtain $T_\mathrm{e} = 29$ $\mu$K (31 $\mu$K) at point C for $\dot{Q} = 1$ nW (2 nW). These estimated electron temperatures are substantially lower than the temperature measured by the Pt NMR thermometer at point C (51 $\mu$K). This discrepancy strongly suggests the presence of a small local heat leak directly into the Pt NMR thermometer, causing only the thermometer temperature to be higher than $T_\mathrm{e}$. Indeed, in a separate cooling run, $T_\mathrm{Pt}$ decreased from ∼100 to

~50 μK when a resistor (~3 Ω) was connected in parallel with the Pt NMR magnet. This resistor and the magnet inductance formed a low-pass filter, suppressing low-frequency noise from the current source and thereby reducing eddy-current heating within the thermometer (details are given in Appendix E). Although the data presented in this article were obtained after this improvement, a non-negligible residual heat leak was still present in the Pt NMR thermometer (Appendix E).

We now perform a comprehensive analysis of the temperature evolution in the region shown in Fig. 2(d) by employing numerical simulations. In this analysis, we take into account the field sweep $\dot{B}$ and the associated eddy-current heat leak $\dot{Q}_{\mathrm{eddy}}$. In this case, instead of Eq. (1), we use

$$\frac{dT_{\mathrm{n}}}{dt} = \frac{\dot{Q}+\dot{Q}_{\mathrm{eddy}}}{n\Lambda B^2} T_{\mathrm{n}}^2 + \frac{\dot{B}}{B} T_{\mathrm{n}}, \qquad (5)$$

where the second term represents the effect of the magnetic-field sweep. In the presence of $\dot{Q}_{\mathrm{eddy}}$, Eq. (4) is also replaced by

$$\frac{T_{\mathrm{e}}}{T_{\mathrm{n}}} = 1 + \frac{\kappa(\dot{Q}+\dot{Q}_{\mathrm{eddy}})}{n\Lambda B^2}. \qquad (6)$$

We also take into account a local heat leak to the Pt NMR thermometer, $\dot{Q}_1$, as discussed above. With this heat leak, a temperature difference is established between $T_{\mathrm{Pt}}$ and $T_{\mathrm{e}}$. By integrating the heat-flow equation combined with the Wiedemann–Franz law, we have

$$T_{\mathrm{Pt}}^2 - T_{\mathrm{e}}^2 = \frac{2R_{\mathrm{el}}^{\mathrm{Pt}}}{L_0} \dot{Q}_1, \qquad (7)$$

where $L_0$ is the Lorenz number, $R_{\mathrm{el}}^{\mathrm{Pt}}$ is the electrical resistance between the location of the heat leak in the Pt NMR thermometer and the lower plate of the experimental space (see Fig. 4), and $R_{\mathrm{th}}^{\mathrm{Pt}}$ is the corresponding thermal resistance, given by $R_{\mathrm{th}}^{\mathrm{Pt}} = R_{\mathrm{el}}^{\mathrm{Pt}}/L_0 T$ . Here the thermal resistance between the lower plate and the copper stage has been neglected because the corresponding electrical resistance is estimated to be only about 1 nΩ.

We perform a numerical simulation by solving Eqs. (5)–(7) so that the simulated $T_{\mathrm{Pt}}$ reproduces the experimental temperatures at the initial and final points around the minimum field [points A and D in Fig. 2(d)], as well as the temperature evolution during the hold period at the minimum field (between points B and C). Details of the numerical-simulation procedure are given in Appendix C. In the simulation, we assume that the background heat leak $\dot{Q}$ is constant and use $\dot{Q}_{\mathrm{eddy}} = 4.1$ nW for the sweep rate employed in the temperature region shown in Fig. 2(d) ($\dot{B} = 1.2\times10^{-5}$ T/s). This value of $\dot{Q}_{\mathrm{eddy}}$ is estimated from the geometry of the copper stage and its RRR [52]. Here, we neglect the magnetoresistance of copper because the resistivity increase is small at the relevant low fields, with $\Delta\rho/\rho_0 \lesssim 0.1$ at 12 mT and $\sim 1.2$ at 100 mT [83], where $\Delta\rho$ is the resistivity increase and $\rho_0$ is the zero-field resistivity. We note that $\dot{Q}_{\mathrm{eddy}}$ has only a minor effect on the simulation results because it occurs only during magnetic-field sweeps, which are much shorter than the constant-field periods.

The results of the simulation are shown in Fig. 2(d). The simulation reproduces the time evolution of $T_{\mathrm{Pt}}$ over the entire time range very well. In the figure, $T_{\mathrm{n}}$ and $T_{\mathrm{e}}$ are also displayed. Just after the minimum field is reached, $T_{\mathrm{n}}$ is estimated to reach as low as 18 $\mu$K and then increase to 26 $\mu$K after 37 hours. According to Eq. (6), the electron temperature $T_{\mathrm{e}}$ of the stage is only 17% higher than $T_{\mathrm{n}}$. In contrast, a large temperature difference is established between $T_{\mathrm{Pt}}$ and $T_{\mathrm{e}}$. The simulation yields $\dot{Q} = 2.2$ nW and $R_{\mathrm{el}}^{\mathrm{Pt}}\dot{Q}_1 = 2.1\times10^{-17}$ WΩ. The former value is in reasonable agreement with the heat leak of 1–2 nW evaluated from the increase in temperature observed at a constant magnetic field (Fig. 3). The magnitude of $R_{\mathrm{el}}^{\mathrm{Pt}}\dot{Q}_1$ quantifies the temperature difference between $T_{\mathrm{Pt}}$ and $T_{\mathrm{e}}$. This value is about two orders of magnitude larger than that expected from a simple estimate based on the residual resistance of the Pt NMR thermometer components and the eddy-current heating caused by current fluctuations in the magnet. (Details are given in Appendix E.) Identifying the exact heat-leak path and eliminating the heat leak are important for accurate measurements of the electron temperature below $\sim$100 $\mu$K.

The slow increase in temperature demonstrated above suggests that the refrigerator can keep temperatures in the deep sub-mK regime for an extended period of time. To estimate how long such temperatures can be sustained, we further perform the following two simulations using the parameters obtained above.

The first simulation is to estimate the holding time below 100 $\mu$K if the magnetic field is maintained at the minimum field after point C in Fig. 2(d), instead of being ramped up subsequently. The results of the simulation are shown in Fig. 5(a). (Details of the simulation procedure are described in Appendix C.) The temperature of the Pt NMR thermometer can remain below 100 $\mu$K for a total of 90 hours after reaching the minimum field. This is because a substantial amount of available entropy is still left in the copper nuclear-spin system at point C.

The second simulation is to estimate the holding time at $T_{\mathrm{Pt}} = 100\ \mu\mathrm{K}$ by slowly demagnetizing the stage to compensate for the heat leak. This simulation is also performed by solving Eqs. (5)–(7) under the condition of constant $T_{\mathrm{Pt}}$. (The detailed procedure is described in Appendix D.) Figure 5(b) shows the results of this simulation, in which the demagnetization is stopped when $T_{\mathrm{Pt}}$ reaches $100\ \mu\mathrm{K}$ during the demagnetization shown in Fig. 2(d). All temperatures ($T_{\mathrm{Pt}}$, $T_{\mathrm{e}}$, and $T_{\mathrm{n}}$) are nearly constant for a long period until the magnetic field decreases to a very small value (∼20 mT). The temperature of the Pt NMR thermometer stays at 100 $\mu$K for a remarkably long period of more than 300 hours. In practice, although the magnetic-field sweep cannot be controlled exactly as in the simulation, a comparable holding time should be achievable by following a field-sweep profile close to the simulated one.

The above experimental results, together with these two simulations, demonstrate that the refrigerator has sufficient performance to conduct long-duration experiments in the deep sub-mK regime, even for samples with thermal relaxation times as long as several days.

### B. Demagnetization after short precooling

The small heat leak demonstrated in Sect. III A allows the refrigerator to reach temperatures well below 100 $\mu$K even with a short precooling time. To demonstrate this, we precool the stage at 9 T for 12 hours down to 19.4 mK and then initiate demagnetization, as shown in Fig. 6(a). In this case, the reduction of the entropy of the nuclear-spin system is only 3.0%. After opening the heat switch, the magnetic field is swept down to the minimum field of 12 mT over 28 hours, reaching a temperature of 59 $\mu$K as measured by the Pt NMR thermometer [Fig. 6(b)]. The magnetic field is then kept at this value for 32 hours, during which the temperature increases from 59 to 73 $\mu$K [Fig. 6(c)].

We perform a detailed analysis of the temperature evolution in the vicinity of the minimum field by employing a numerical simulation similar to that used for the demagnetization run in Sect. III A. As shown in Fig. 6(c), the simulation reproduces the time evolution of $T_{\mathrm{Pt}}$. The simulation shows that the nuclear-spin temperature reaches 30 $\mu$K just after the field is reduced to 12 mT and increases to 61 $\mu$K after a 32-hour holding period. The obtained values, $\dot{Q} = 2.3$ nW and $R_{\mathrm{el}}^{\mathrm{Pt}}\dot{Q}_1 = 2.1 \times 10^{-17}\ \mathrm{W\,\Omega}$, are very close to those obtained for the demagnetization run in Fig. 2. These results support the same interpretation as in Sect. III A:

during this period, $T_{\mathrm{n}}$ and $T_{\mathrm{e}}$ remain at substantially lower temperatures, whereas only the Pt NMR thermometer is heated by a small local heat leak.

To estimate how long the deep sub-mK regime can be sustained, we perform two numerical simulations similar to those for the demagnetization run described in Sect. III A. The first simulation evaluates how long $T_{\mathrm{Pt}}$ remains below 100 $\mu$K if the magnetic field is kept at the minimum field after point C instead of being increased toward point D. As shown in Fig. 7(a), $T_{\mathrm{Pt}}$ remains below 100 $\mu$K for more than 39 hours after reaching the minimum field. The second simulation evaluates the holding time for keeping $T_{\mathrm{Pt}}$ at 100 $\mu$K by slowly demagnetizing the stage to compensate for the heat leak. The simulation shows that the refrigerator can keep $T_{\mathrm{Pt}}$ at 100 $\mu$K for ~87 hours, as displayed in Fig. 7(b).

The above experimental results, together with the simulation results, indicate that even after a short precooling time of 12 hours, the refrigerator provides a sufficiently long measurement time in the deep sub-mK regime for samples with thermal relaxation times that are not excessively long.

## IV. CONCLUSIONS

We demonstrate that the nuclear demagnetization refrigerator provides efficient access to the deep sub-mK regime. A full performance characterization shows that the temperature of the Pt NMR thermometer reaches 50 $\mu$K and that the refrigerator can maintain a temperature of 100 $\mu$K for more than ten days. Even after a short precooling time, the refrigerator sustains temperatures in the deep sub-mK regime for much longer than the precooling time itself, allowing most of the available time to be devoted to sample measurements. This feature also enables the precooling time to be adjusted according to experimental requirements. Together with its versatile experimental design, this capability enables rapid and efficient access to the deep sub-mK regime for a wide range of studies in condensed matter systems, including studies of quantum devices and quantum materials.

In addition, this study employs heat-flow modeling and numerical simulations to identify the problem of a local heat leak to the Pt NMR thermometer. We show that the modeling and simulations also provide estimates of the achievable temperature and holding time. This approach provides an effective way to identify sources of unwanted heat leaks and other thermal problems, while enhancing the efficiency of planning experiments in the deep sub-mK regime.

**AUTHOR CONTRIBUTIONS**

H.I. conceived the idea and supervised the project. Z.L., J.S., and H.I. designed the system. All authors participated in the construction of the adiabatic nuclear demagnetization refrigerator. Z.L. performed the experiments and data analysis with the help of J.S. and B.T. H.I. wrote the manuscript with input from all authors.

**DECLARATION OF COMPETING INTERESTS**

The authors declare that they have no known competing financial interests or personal relationships that could have appeared to influence the work reported in this paper.

**ACKNOWLEDGMENTS**

This work was partially supported by the National Natural Science Foundation of China (Grant Nos. 12474135, 92365207), the National Basic Research Program of China (Grant No. 2025YFA1411400), and the Synergetic Extreme Condition User Facility (SECUF).

**Appendix A: Diffusion bonding of aluminum and copper for the heat switch**

We use aluminum foils with a thickness of 0.25 mm (purity of 99.9995%). The foils are annealed in advance at 500 °C in a vacuum for 3 hours. The resulting RRR is 4300. For copper, plates with a thickness of 1 mm (purity of 99.999%) are annealed at 900 °C in an oxygen atmosphere at $5\times10^{-5}$ mbar for 24 hours. The resulting RRR is 5800. Both materials are acid-etched immediately before diffusion bonding. For aluminum, a mixed acid of 80% $H_3PO_4$, 5% $HNO_3$, 5% $CH_3COOH$, and 10% $H_2O$ is used for 5 minutes. For copper, 5% $HNO_3$ is used for 5 minutes.

The heat switch is composed of four aluminum foils sandwiched alternately between a total of five copper plates [Fig. 1(b)]. Each aluminum foil is in contact with copper plates on both sides, with a contact area of $5\times8$ mm$^2$ on each side. The sandwiched assembly is placed between two stainless-steel jigs with molybdenum foils inserted on both sides. This assembly is then pressed using an oil piston cylinder at a pressure of 100 MPa. After 10 minutes of pressing, the jigs are tightly fastened with two M6 titanium screws, and then the entire assembly is heat-treated in a vacuum ($\sim 5\times10^{-6}$ mbar). From our test experiments [78], we found that heat treatment at 500 °C for 180 minutes is optimal. Under these conditions, the contact resistance measured at 4.2 K is 81 nΩ for each aluminum–copper contact (contact area of $5\times8$ mm$^2$). From this value, the total resistance between the two flanges on either side of the heat switch is estimated to be 27 nΩ. We note that this value is one order of magnitude larger than that of a state-of-the-art aluminum–copper-joint heat switch fabricated by plasma-etching the aluminum oxide layer and then depositing gold without breaking the vacuum of the e-beam evaporator [84]. If such equipment is available, this fabrication method could provide a promising alternative approach.

**Appendix B: Numerical simulations of precooling of the copper stage**

We perform numerical simulations of precooling curves. For this purpose, we use the heat path model shown in Fig. 8(a). Heat from the copper stage flows through the thermal resistance $R_{\rm th}$ to the mixing chamber, where it is removed with a cooling power $P(T_{\rm mc})$. Here $T_{\rm mc}$ is the temperature of the mixing chamber plate. Consequently, the temperature of the nuclear-spin system, $T_{\rm n}$, decreases according to

$$\begin{aligned} P(T_{\rm mc}) &= -C_{\rm n}\dot{T}_{\rm n} \\ &= -n\Lambda\frac{B^2}{T_{\rm n}^2}\frac{dT_{\rm n}}{dt}, \end{aligned} \qquad \text{(B1)}$$

where $C_{\rm n}$ is the heat capacity of the nuclear-spin system, and the second equality follows from the high-temperature approximation [Eq. (2)].

Heat flows from the stage to the mixing chamber through the thermal resistance $R_{\rm th}$, resulting in a temperature difference between $T_{\rm mc}$ and $T_{\rm n}$, where $R_{\rm th}$ is the total thermal resistance between the mixing chamber plate and the copper stage. Assuming the Wiedemann–Franz law, $R_{\rm th}$ at temperature $T$ is related to the electrical resistance $R_{\rm el}$ by $R_{\rm th} = R_{\rm el}/L_0T$, where $L_0$ is the Lorenz number. For a small temperature difference $\Delta T$, the temperature drop across the thermal resistance is given by $\Delta T = \frac{R_{\rm el}}{L_0T}P$. In our case, the temperature difference between $T_{\rm mc}$ and $T_{\rm n}$ is finite. Therefore, we use the integrated form of this equation,

$$T_{\rm n}^2 - T_{\rm mc}^2 = \frac{2R_{\rm el}}{L_0} P(T_{\rm mc}). \qquad \text{(B2)}$$

We numerically solve Eqs. (B1) and (B2) to obtain $T_{\rm n}$ and $T_{\rm mc}$ as functions of time. The calculations are performed at $B = 9$ T using the cooling power, $P(T_{\rm mc})$, measured at the mixing chamber plate, as shown in Fig. 8(b). The calculated precooling curves are shown in Fig. 8(c) as functions of time for different values of $R_{\rm el}$. As $R_{\rm el}$ increases, the precooling process becomes progressively slower. The precooling time required for $T_{\rm n}$ to decrease from 50 mK to 11.5 mK [the precooling endpoint in Fig. 2(a)] is 66, 73, 80, 102, and 139 hours for $R_{\rm el} = 0$, 100, 200, 500, and 1000 nΩ, respectively.

**Appendix C: Numerical simulations of the temperature evolution during demagnetization**

In the main text, we presented numerical analyses of the behavior near the minimum magnetic field in Figs. 2(d) and 6(c). In this section, we describe the procedure used in these analyses in detail.

The simulations are based on Eqs. (5), (6), and (7), which are reproduced below:

$$\frac{dT_{\rm n}}{dt} = \frac{\dot{Q}+\dot{Q}_{\rm eddy}}{n\Lambda B^2} T_{\rm n}^2 + \frac{\dot{B}}{B} T_{\rm n}, \qquad \text{(C1)}$$

$$\frac{T_{\rm e}}{T_{\rm n}} = 1 + \frac{\kappa(\dot{Q}+\dot{Q}_{\rm eddy})}{n\Lambda B^2}, \qquad \text{(C2)}$$

and

$$T_{\rm Pt}^2 - T_{\rm e}^2 = \frac{2R_{\rm el}^{\rm Pt}}{L_0} \dot{Q}_1 \equiv A. \qquad \text{(C3)}$$

Equation (C1) describes the time evolution of $T_{\rm n}$ caused by the total heat leak to the stage $\dot{Q}$ and the eddy-current heating $\dot{Q}_{\rm eddy}$, as well as the adiabatic process associated with the magnetic-field sweep $\dot{B}$. Equation (C2) describes the increase of $T_{\rm e}$ above $T_{\rm n}$ caused by $\dot{Q}$ and $\dot{Q}_{\rm eddy}$. Equation (C3) describes the temperature difference between $T_{\rm Pt}$ and $T_{\rm e}$ arising from the local heat leak to the Pt NMR thermometer $\dot{Q}_1$. As noted in the main text, $\dot{Q}_{\rm eddy}$ has only a minor effect on the simulation results, because it is present only during the short magnetic-field sweeps. Below, we use $T_{\rm Pt}^{\rm sim}$ and $T_{\rm Pt}^{\rm exp}$ to denote the temperature of the Pt NMR thermometer obtained from the simulations and the experiments, respectively, to avoid confusion.

In the simulations, Eqs. (C1)–(C3) are solved iteratively assuming a constant $\dot{Q}$. The two unknown parameters, $\dot{Q}$ and $A$, are determined by requiring that $T_{\rm Pt}^{\rm sim}$ reproduces the corresponding experimental values ($T_{\rm Pt}^{\rm exp}$) at the initial and final points [points A and D for the case of Fig. 2(d)], as well as the temperature evolution at the minimum field (between points B and C). We note that $\dot{B}$ and $\dot{Q}_{\rm eddy}$ are known parameters. The detailed procedure is described below for the case in Fig. 2(d).

Step 1:

We set initial trial values for $T_{\rm n}$ at points A and D in Fig. 2(d).

Step 2:

We set an initial guess for $\dot{Q}$. The time evolution of $T_{\rm n}$ is then calculated from Eq. (C1) for the trial value for $T_{\rm n}$ at point A. This calculation is repeated iteratively by adjusting $\dot{Q}$ until the calculated value of $T_{\rm n}$ at point D agrees with the trial value specified for point D. We note that for a constant $B$, Eq. (C1) has the analytical solution,

$$T_{\rm n}(t)^{-1} = T_{\rm n}(0)^{-1} - \frac{\dot{Q}}{n\Lambda B^2}t, \qquad \text{(C4)}$$

which makes the calculations somewhat simpler.

Step 3:

Substituting the value of $\dot{Q}$ obtained in Step 2 into Eq. (C2), the time evolution of $T_{\rm e}$ is calculated over the entire time range shown in Fig. 2(d).

Step 4:

Using the value of $T_{\rm e}$ obtained in Step 3, the parameter $A$ is obtained from Eq. (C3) as the average value of $(T_{\rm Pt}^{\rm exp})^2 - T_{\rm e}^2$ during the minimum-field holding period between points B and C.

Step 5:

With the value of $A$ obtained in Step 4, $T_{\rm Pt}^{\rm sim}$ is calculated over the entire time range shown in Fig. 2(d). If the calculated values of $T_{\rm Pt}^{\rm sim}$ at points A and D are different from the corresponding experimental values ($T_{\rm Pt}^{\rm exp}$), Steps 2-4 are repeated with revised trial values of $T_{\rm n}$ at points A and D until $T_{\rm Pt}^{\rm sim}$ reproduces $T_{\rm Pt}^{\rm exp}$ at these points.

From this procedure, $\dot{Q}$ and $A$, as well as the time evolutions of $T_{\rm n}$, $T_{\rm e}$, and $T_{\rm Pt}^{\rm sim}$ are obtained. In the simulation, $\dot{B} = 1.2 \times 10^{-5}$ T/s and $\dot{Q}_{\rm eddy} = 4.1$ nW (the value estimated from the geometry of the copper stage and its RRR [52]) are used. The obtained parameters are $\dot{Q} = 2.2$ nW and $R_{\rm el}^{\rm Pt}\dot{Q}_1 = AL_0/2 = 2.1{\times}10^{-17}$ WΩ, and the simulation results are shown in Fig. 2(d).

Using the obtained parameters, we further performed a numerical simulation for the case in which the magnetic field is maintained at the minimum field without increasing it after point C, discussed in the main text. The results are shown in Fig. 5(a).

The same procedure is applied to the simulation shown in Fig. 6(c). Points A–D in Fig. 6(c) have the same meanings as those in Fig. 2(d).

**Appendix D: Numerical simulations of slow demagnetization for heat-leak compensation**

In Sects. III A and III B, we discuss the results of the simulations in which the stage is slowly demagnetized to keep $T_{\mathrm{Pt}}$ constant by compensating for the temperature increase arising from the heat leak. Here, we describe the simulation procedure in detail. According to Eq. (C3), maintaining $T_{\mathrm{Pt}}$ constant implies that $T_{\mathrm{e}}$ is also constant. Under this condition, Eqs. (C1) and (C2) yield

$$\frac{dB}{dt} = -\frac{\dot{Q}T_{\mathrm{e}}}{n\Lambda}\frac{B}{\left(B^2 - B_{\mathrm{opt}}^2\right)}, \quad \text{(D1)}$$

where $B_{\mathrm{opt}} = \sqrt{\kappa\dot{Q}/n\Lambda}$. We note that $B_{\mathrm{opt}}$ is the so-called optimal final demagnetization field (see Ref. [52]). In the above, $\dot{Q}_{\mathrm{eddy}}$ is neglected because it is much smaller than $\dot{Q}$ for most of the simulated time owing to the very slow field sweep, except at the very end of the simulation. Equation (D1) has an analytical solution

$$\frac{B^2 - B_0^2}{2} - B_{\mathrm{opt}}^2 \ln\left(\frac{B}{B_0}\right) = -\frac{\dot{Q}T_{\mathrm{e}}}{n\Lambda}(t - t_0), \qquad \text{(D2)}$$

where $B_0$ is the value of $B$ at $t = t_0$.

For the demagnetization run shown in Fig. 2, Eq. (D2) is evaluated for the case in which the demagnetization is stopped when $T_{\mathrm{Pt}}$ reaches 100 $\mu$K during the demagnetization shown in Fig. 2(d). In this simulation, the parameters obtained from the simulations described in Appendix C ($\dot{Q} = 2.2$ nW and $R_{\mathrm{el}}^{\mathrm{Pt}}\dot{Q}_1 = 2.1\times10^{-17}$ WΩ) are used. The simulation results are shown in Fig. 5(b). We note that $T_{\mathrm{e}} = 2T_{\mathrm{n}}$ when $B$ reaches $B_{\mathrm{opt}} = 4.9$ mT at the end of the simulation (311 hours), beyond which $T_{\mathrm{e}}$, and thus $T_{\mathrm{Pt}}$, can no longer be maintained constant.

The same procedure is applied to the demagnetization run shown in Fig. 6.

**Appendix E: Local heat leak to the Pt NMR thermometer**

In a separate cooling run performed prior to the demagnetization runs presented in the main text, we found that the temperature of the Pt NMR thermometer decreased from $\sim 100$ to $\sim 50\ \mu$K when a resistor (~3 Ω) was connected in parallel with the magnet of the Pt NMR thermometer. This indicates the presence of a local heat leak directly entering the Pt NMR thermometer before the resistor was connected. Applying the same analysis as described in the main text, we obtained $R_{\rm el}^{\rm Pt}\dot{Q}_1 = 1.1 \times 10^{-16}$ W Ω before the resistor was connected. It decreased to $2.1{\times}10^{-17}$ WΩ after the resistor was connected. Because $R_{\rm el}^{\rm Pt}$ should not change, the direct heat leak to the Pt NMR thermometer, $\dot{Q}_1$, decreased to about 20% of its original value. This indicates that a non-negligible heat leak still remained even after the resistor was connected.

The resistor, together with the inductance of the magnet (~15 mH measured at 48 Hz), forms a low-pass filter with a cut-off frequency of ~36 Hz, thereby suppressing low-frequency noise generated by the current source in the magnet leads. Although the noise spectrum is unknown, 50 Hz is a plausible frequency for the dominant noise component. Assuming that the noise consists only of a 50 Hz component and causes eddy-current heating within the thermometer, the filter is expected to reduce the resulting heat leak to approximately 30% of its original value, as estimated from the circuit parameters. This value is reasonably consistent with the reduction of $\dot{Q}_1$ to 20% of its original value obtained above, considering that noise components at frequencies other than 50 Hz may also be present.

In contrast, no apparent reduction in the total heat leak to the stage (typically 1–2 nW) was observed after connecting the resistor, indicating that $\dot{Q}_1$ is on the order of 0.1 nW or less. This heat leak could likely be further reduced by using a current source with lower low-frequency noise, which may allow the Pt NMR thermometer to be used at even lower temperatures. Indeed, temperatures as low as 12 and 27 $\mu$K have been measured using Pt NMR thermometers in previous works [58,60]. We note that noise thermometry provides an alternative method for measuring temperatures down to $\sim 50\ \mu$K and below, with a heat input as low as 1 fW [85].

We now consider the implications of the obtained value of $R_{\rm el}^{\rm Pt}\dot{Q}_1$. To this end, we estimate the eddy-current heating arising from magnet-current fluctuations and the thermal resistance of the components within the Pt NMR thermometer. Although the actual noise frequency and amplitude are unknown, we assume that the current noise has a frequency of 50 Hz and an amplitude of 6.5 $\mu$A. The latter was obtained by scaling the current ripple given in the specifications (30 $\mu$A at an output current of 2.5 A) of the current source to the operating current of the magnet (0.54 A). (The current source used was the current supply module of PLM-5, CS-10.) The magnetic-field fluctuation caused by this current ripple is $\delta B = 2.4\times10^{-4}$ mT, which generates eddy-current heating in the thermometer. Our estimate indicates that the silver support for the Pt wire assembly makes the largest contribution to $R_{\mathrm{el}}^{\mathrm{Pt}}\dot{Q}_1$ of all the components. From our estimate, the eddy-current heating in the support is about 0.04 nW based on its geometry, the RRR (= 1000) of the material, and the magnetic-field profile. The residual resistance between the top of the support and its bottom, which is screwed to the lower plate of the experimental space, is about 20 nΩ. The contact resistance at the lower plate is about 10 nΩ. Thus, $R_{\mathrm{el}}^{\mathrm{Pt}}\dot{Q}_1$ for the support is about $1\times10^{-18}$ WΩ, which is about two orders of magnitude smaller than the experimentally obtained value ($1.1\times10^{-16}$ WΩ). This value is further reduced to about $3\times10^{-19}$ WΩ when the resistor is connected to form the low-pass filter.

We also consider the Joule heating at the coil associated with the rf pulse. From the pulse amplitude (3.906 $V_{\mathrm{p-p}}$), the pulse length (5.61 $\mu$s) used near the lowest temperature, and the parameters of the NMR measurement circuit, the heating at the coil is estimated to be about 0.2 nJ per pulse. At the lowest temperature, a pulse is applied every 110 min, corresponding to an average heat leak of 30 fW. Because the coil is thermally anchored to the silver support, this heat flows to the support. The resulting contribution to $R_{\mathrm{el}}^{\mathrm{Pt}}\dot{Q}_1$ is less than $1\times10^{-21}$ WΩ, much smaller than the value obtained experimentally above ($2.1\times10^{-17}$ WΩ). To further rule out this possibility, we varied the pulse interval between $5T_1$ and $10T_1$ around the lowest temperature and found no difference in the detected signal, confirming that pulse-induced heating is negligible, where $T_1$ is the spin-lattice relaxation time estimated from the Korringa relation $T_1T = 0.030$ Ks [52].

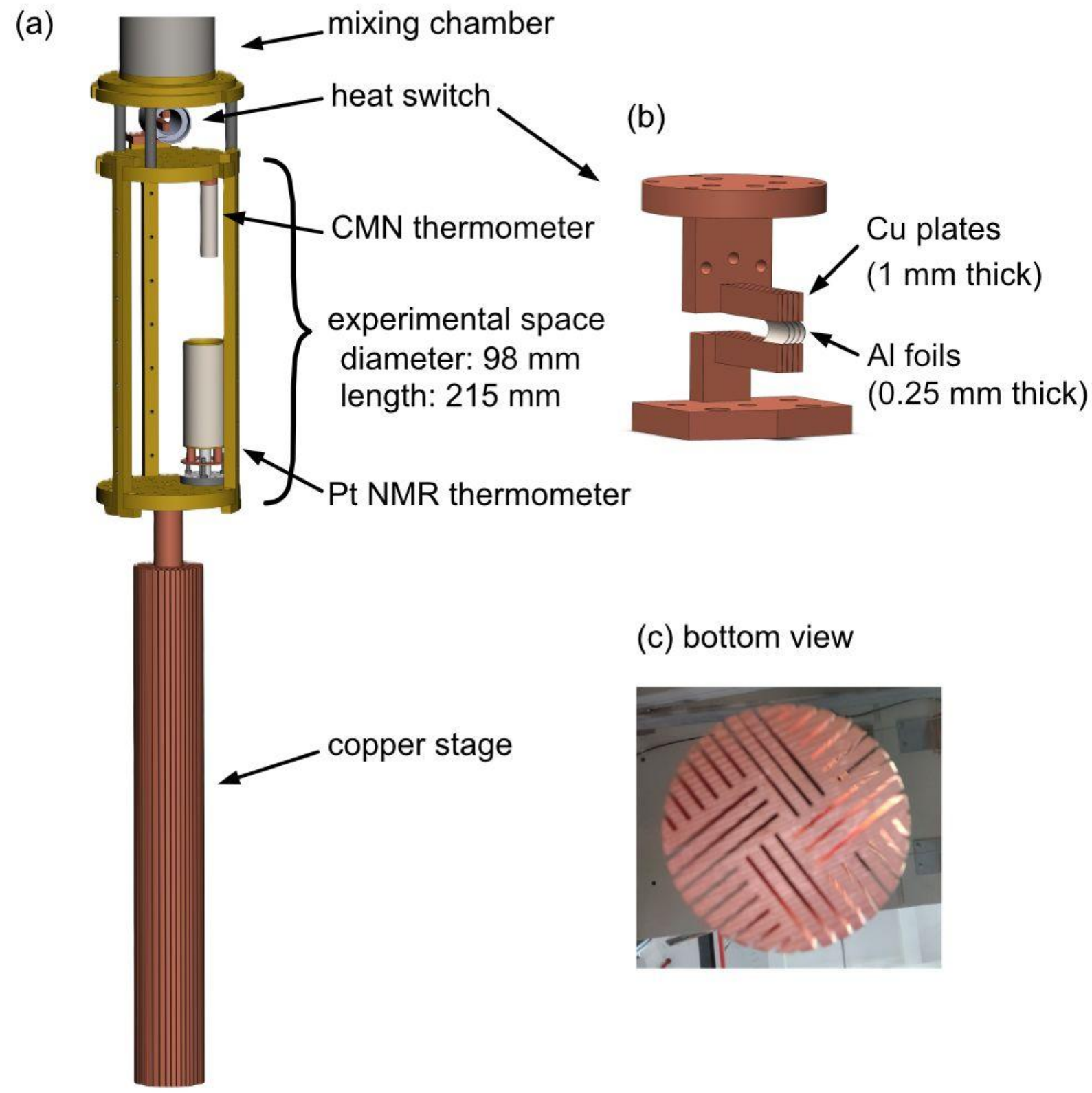

**FIG. 1.** The adiabatic nuclear demagnetization system. (a) Schematic illustration of the system below the mixing chamber inside the IVC. (b) Illustration of the heat switch. (The magnet is not shown.) Four aluminum foils are diffusion bonded to copper plates. (c) Photo of the copper stage (bottom view). The stage has 0.8-mm-wide slits to reduce eddy-current heating.

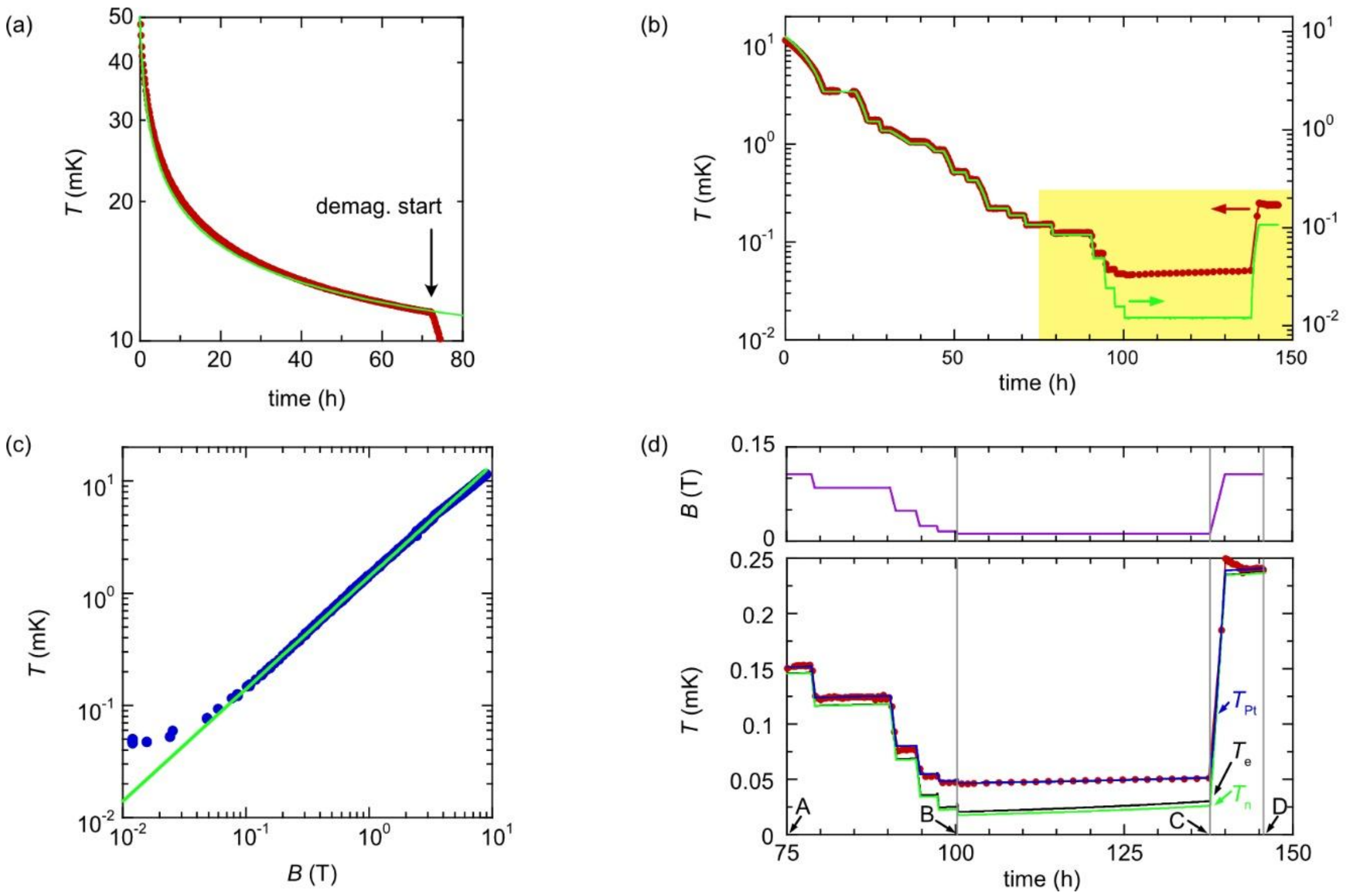

**FIG. 2.** Performance measurement of the adiabatic nuclear demagnetization. (a) Precooling curve of the copper stage at 9 T. The red dots represent experimental data. The green line shows the simulated precooling curve for $R_{\mathrm{el}} = 130$ nΩ (see text). (b) $T$ and $B$ as functions of time after the start of demagnetization. The right axis is scaled relative to the left axis such that $B/T$ is constant, with a scaling factor of 2.43 T/3.41 mK. (c) $T$ as a function of $B$ during cooling. The solid line represents ideal adiabatic behavior [$B/T =$ const. (= 2.43 T/3.41 mK)]. (d) Enlarged view of the yellow region in (b) on a linear scale. The red points in the lower panel represent experimental data, while the blue, black, and green lines represent the simulated $T_{\mathrm{Pt}}$, $T_{\mathrm{e}}$, and $T_{\mathrm{n}}$, respectively.

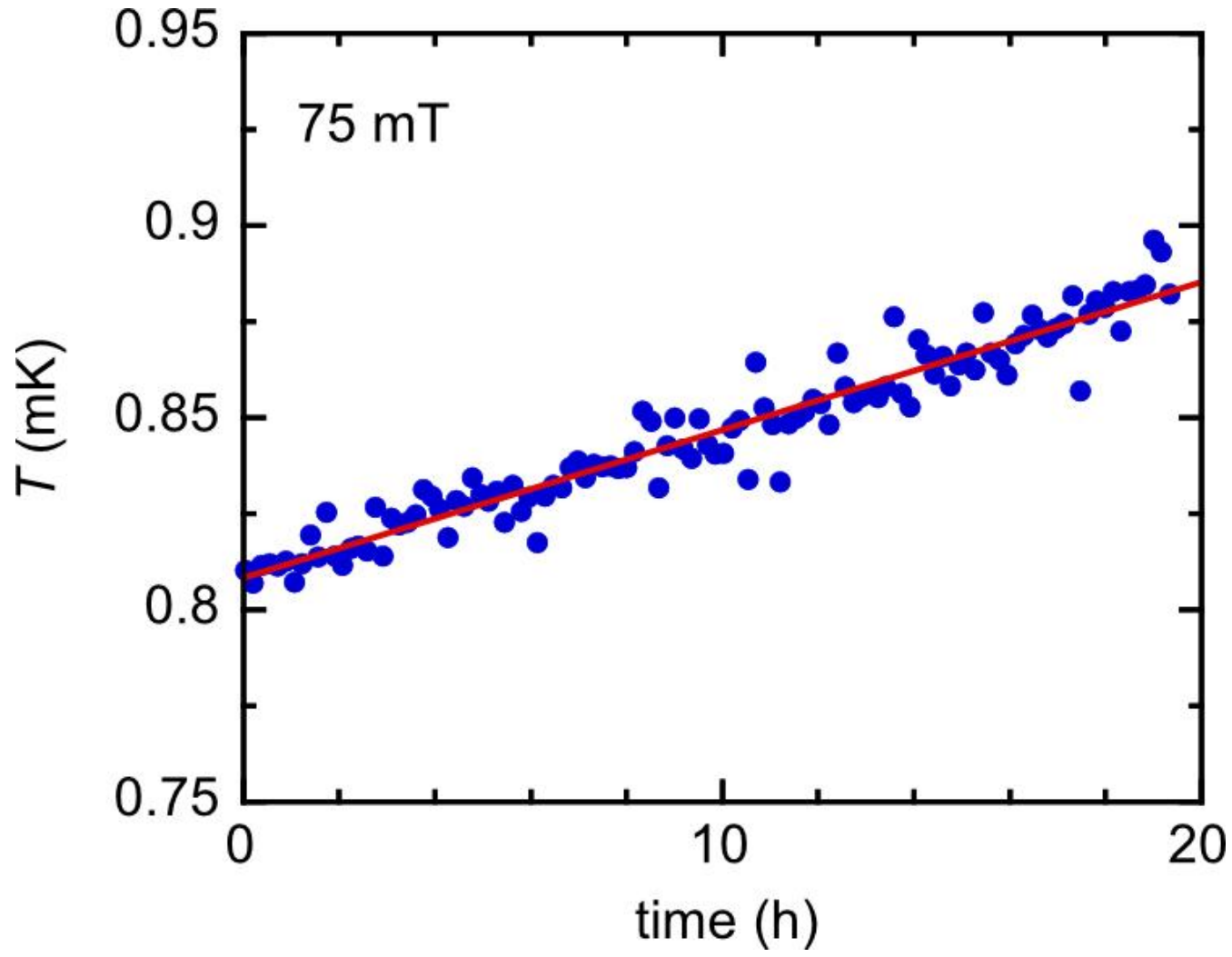

**FIG. 3.** Evaluation of the heat leak to the copper stage. The increase in temperature is monitored while the magnetic field is kept constant (75 mT). The solid line is a linear fit as a function of time. The rate of temperature increase is $\dot{T} = 1.1 \times 10^{-9}$ K/s and the heat capacity is $C_\mathrm{n} \sim 0.88$ J/K, yielding a heat leak of $\dot{Q} \sim 1.0$ nW.

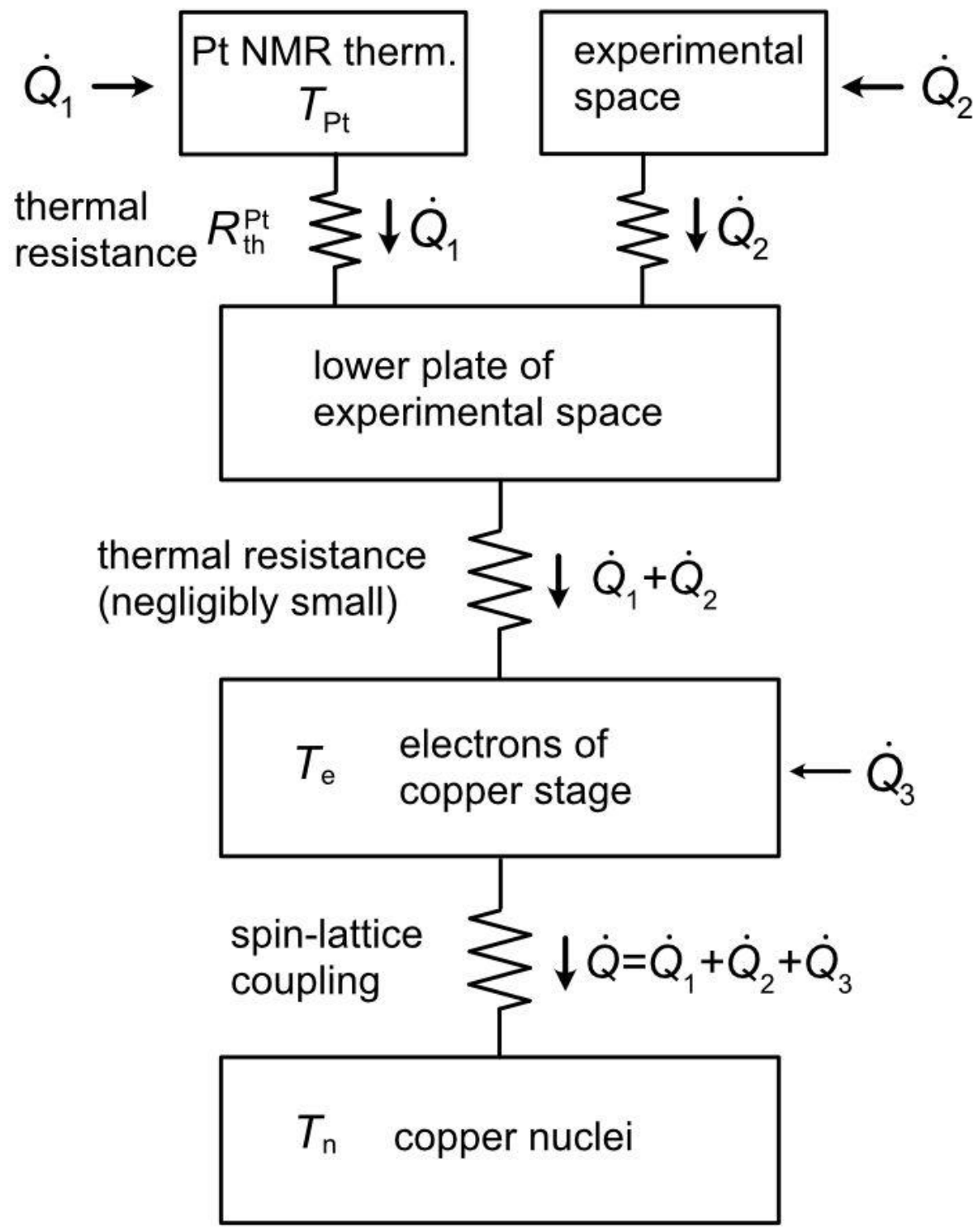

**FIG. 4.** Schematic model of heat flow. All heat leaks first flow into the electron system in the copper stage and are subsequently transferred to the copper nuclei through spin–lattice coupling. The thermal resistance between the lower plate of the experimental space and the stage is estimated to be ~ 1 nΩ and therefore negligible.

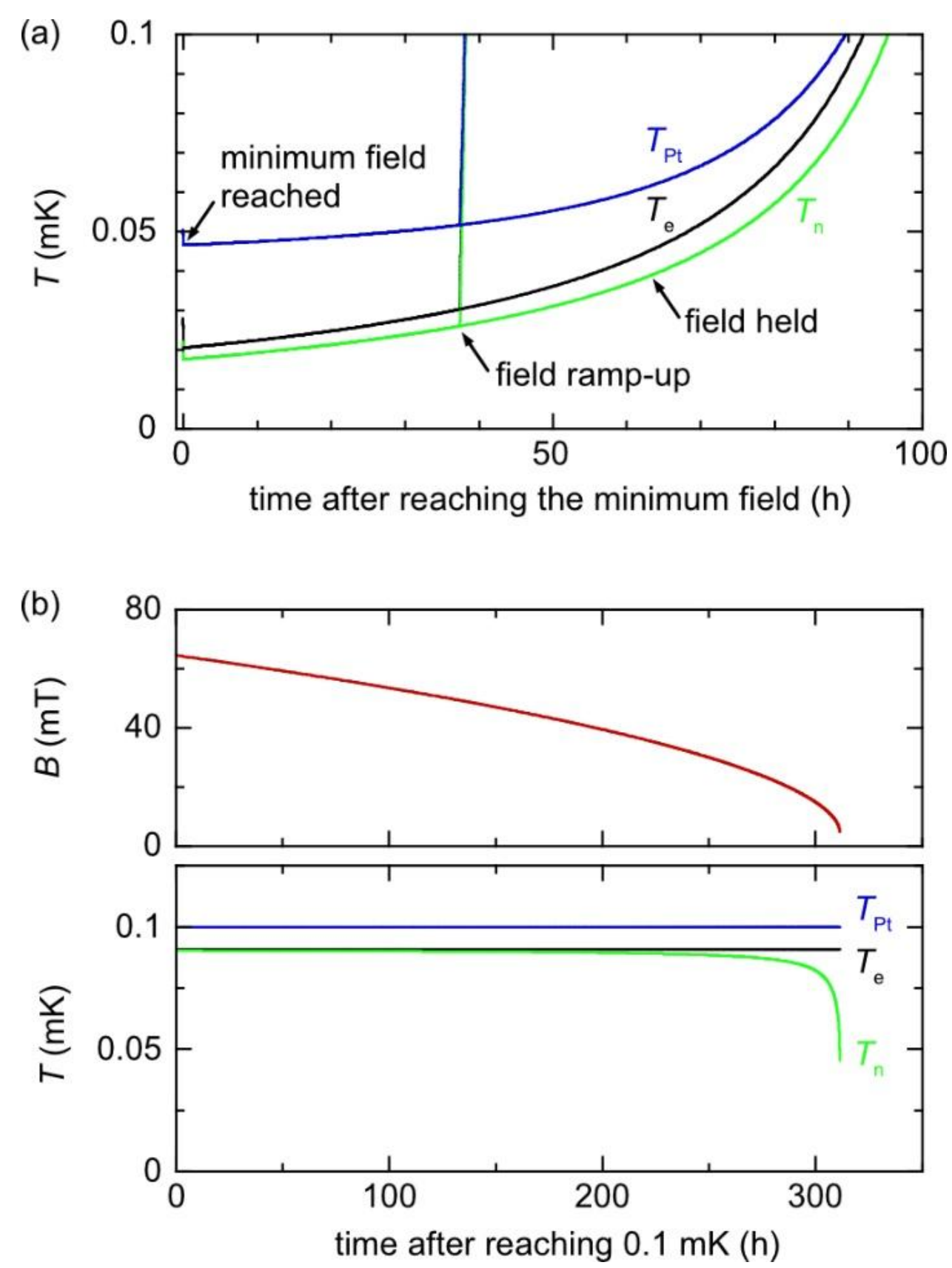

**FIG. 5.** Numerical simulations for estimating the holding time in the deep sub-mK regime for the demagnetization run shown in Fig. 2. (a) Simulated temperatures assuming that the magnetic field is maintained at the minimum field (12 mT) after point C in Fig. 2(d), instead of being ramped up. The 'field ramp-up' indicated in the figure corresponds to point C. (b) Simulated temperatures (lower panel) and magnetic field (upper panel) for the case in which $T_{\mathrm{Pt}}$ is maintained at 100 $\mu$K after reaching this temperature.

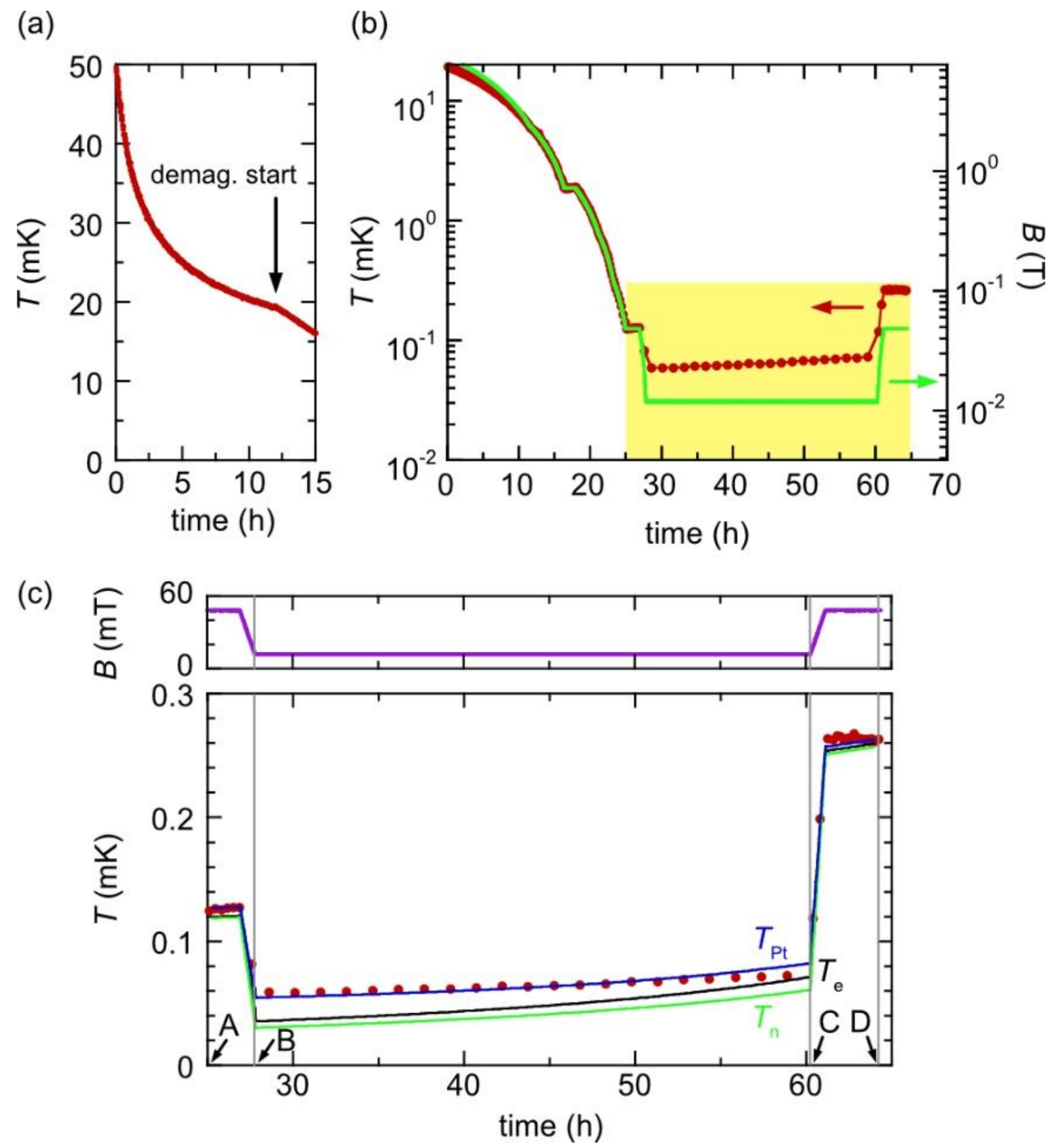

**FIG. 6.** Behavior of demagnetization after short precooling. (a) Precooling curve. Demagnetization is started after 12 hours when the stage reaches 19.4 mK. (b) $T$ and $B$ as functions of time after the start of demagnetization. The right axis is scaled relative to the left axis such that $B/T$ is constant, with a scaling factor of 0.724 T/1.87 mK. (c) Enlarged view of the yellow region in (b) on a linear scale. Red points in the lower panel represent experimental data, while the blue, black, and green lines represent simulated $T_{\mathrm{Pt}}$, $T_{\mathrm{e}}$, and $T_{\mathrm{n}}$, respectively.

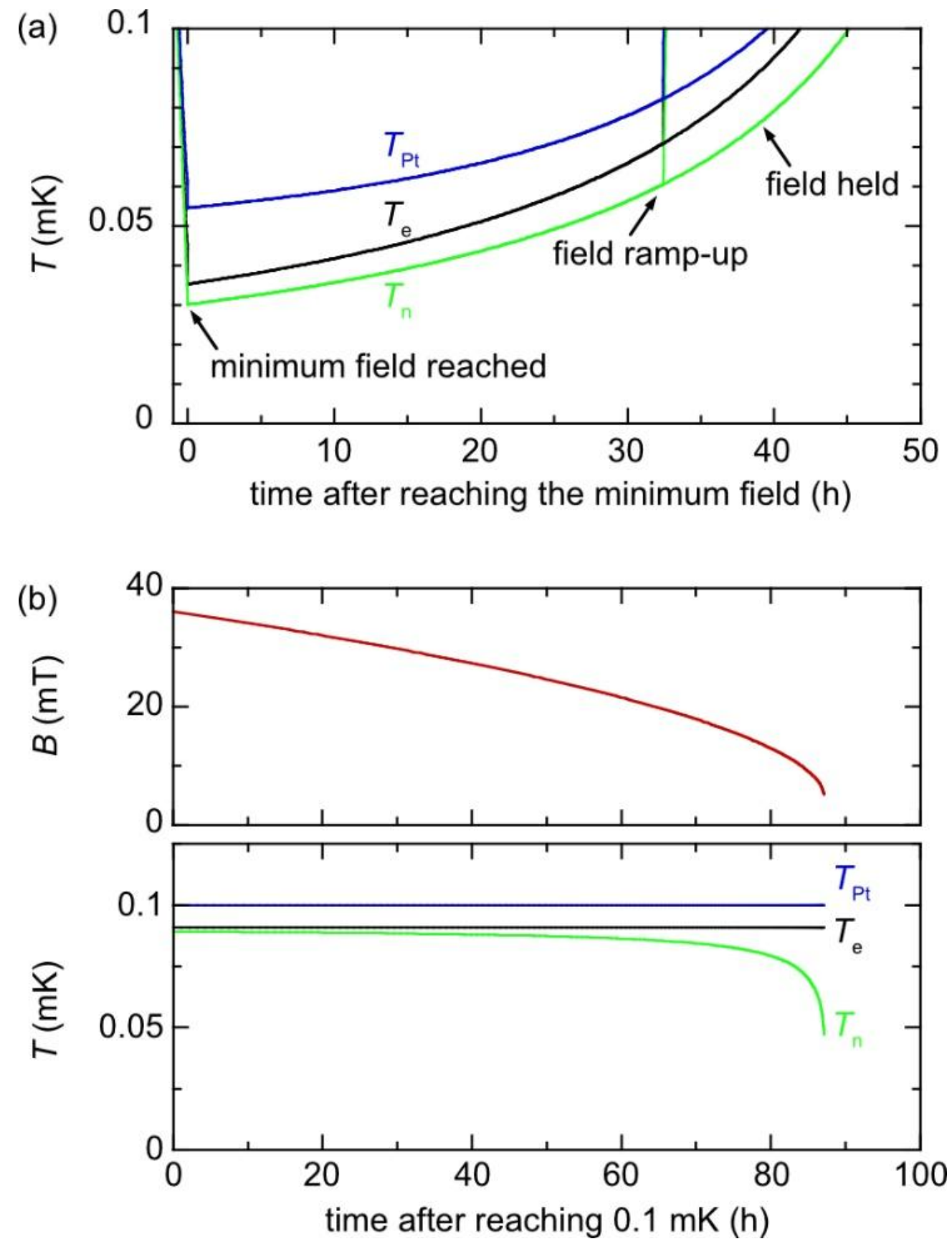

**FIG. 7.** Numerical simulations for estimating the holding time in the deep sub-mK regime for the demagnetization run shown in Fig. 6. (a) Simulated temperatures assuming that the magnetic field is maintained at the minimum field (12 mT) after point C in Fig. 6(c). (b) Simulated temperatures (lower panel) and magnetic field (upper panel) for the case in which $T_{\mathrm{Pt}}$ is maintained at 100 $\mu$K after reaching this temperature.

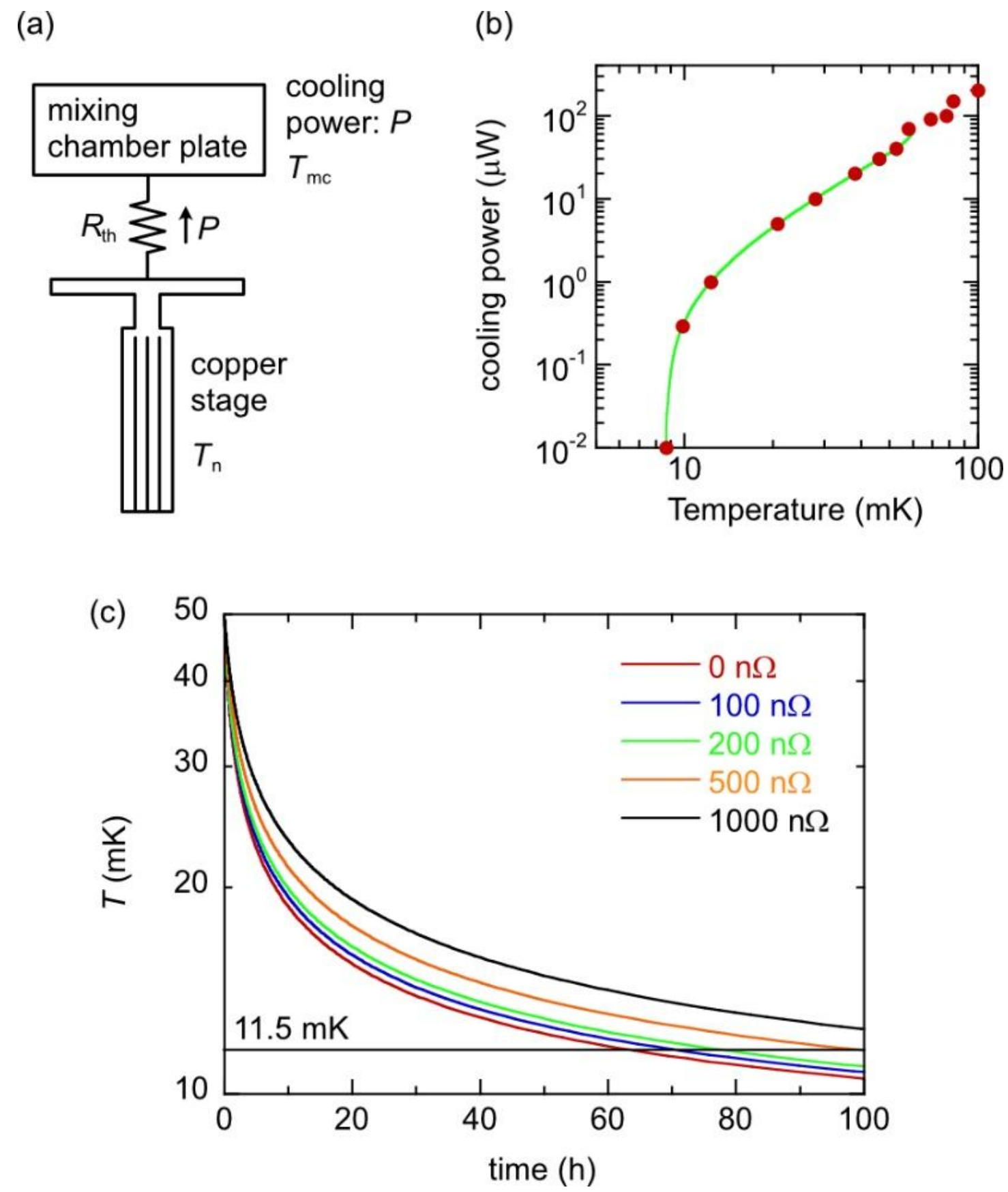

**FIG. 8.** Numerical simulations of the precooling process. (a) Heat flow model used in the simulations. (b) Cooling power measured at the mixing chamber plate under the conditions used for precooling the stage (still-heater power: 10 mW). The solid line represents the fitted curve used in the simulations. The fit is performed below 60 mK. (c) Simulated temperature of the copper stage as a function of time for different values of $R_{\mathrm{el}}$. The horizontal line indicates 11.5 mK, the temperature corresponding to the end of the precooling process in Fig. 2(a).